\documentclass[conference]{IEEEtran}
\usepackage{cite}

\usepackage{placeins}
\usepackage{color}

\usepackage{graphicx}

\usepackage[cmex10]{amsmath}

\usepackage{array}

\usepackage{mdwmath}
\usepackage{mdwtab}

\usepackage[caption=false,font=footnotesize]{subfig}

\usepackage{stfloats}

\usepackage{url}
\IEEEoverridecommandlockouts
\begin{document}

\title{Impact of Dual Slope Path Loss on User Association in HetNets}


\author{\IEEEauthorblockN{$^\star$Nikhil~Garg\thanks{The work was done while the authors were at The University of Texas at Austin.}, $^\dagger$Sarabjot~Singh, and $^\diamond$Jeffrey~Andrews}
\IEEEauthorblockA{$^\star$Department of Electrical Engineering, Stanford University, Palo Alto, CA}
\IEEEauthorblockA{$^\dagger$Intel Labs, Santa Clara, CA}
\IEEEauthorblockA{$^\diamond$Department of Electrical and Computer Engineering, The University of Texas at Austin, Austin, Texas
}
}

\maketitle

\begin{abstract}
Intelligent load balancing is essential to fully realize the benefits of dense heterogeneous networks. Current techniques have largely been studied with single slope path loss models, though multi-slope models are known to more closely match real deployments. This paper develops insight into the performance of biasing and uplink/downlink decoupling for user association in HetNets with dual slope path loss models. It is shown that dual slope path loss models change the tradeoffs inherent in biasing and reduce gains from both biasing and uplink/downlink decoupling. The results show that with the dual slope path loss models, the bias maximizing the median rate is not optimal for other users, e.g., edge users.  Furthermore, optimal downlink biasing is shown to realize most of the gains from downlink-uplink decoupling. Moreover, the user association gains in dense networks are observed  to be quite sensitive to  the path loss exponent beyond the critical distance in a dual slope model.

\end{abstract}

%
\IEEEpeerreviewmaketitle

\section{Introduction} 
Mobile traffic has risen critically in recent years. In 2014 alone, global mobile data traffic grew 69\%, the number of mobile devices grew 7.2\% to 7.4 billion, and average smart phone usage grew 45\%, largely due to the spread of LTE \cite{_cisco_????}. As traffic has increased, it has become beneficial to  offload data to densely deployed pico or femto cells in heterogeneous networks (HetNets). Base stations (BSs) on such tiers, with smaller transmit powers and potentially different propagation characteristics, are cheaper and easier to install than macro BSs. In 2014, 46\% of mobile data traffic was offloaded to Wi-Fi or cellular femto cells, blunting some of the growth of cellular data traffic \cite{_cisco_????}. Optimizing user association and coordination for HetNets is an open and active research area\cite{yiakoumis_behop:_2014, andrews_overview_2014, jo_heterogeneous_2012, damnjanovic_survey_2011, ghosh_heterogeneous_2012,wang_load-aware_2014,ye_user_2013}. In addition, the uplink is considered increasingly important for anticipated applications, and uplink user association has recently been given attention\cite{smiljkovikj_analysis_2015,elshaer_downlink_2014, singh_joint_2015, boccardi_why_2015}. 

\subsection{Background and Related Work}

Many of the benefits of dense networks may not be realized if enough users do not connect to the small-cell BSs.  In \cite{yiakoumis_behop:_2014}, it was observed through a test bed of a dense Wi-Fi network that clients often connected to a 2.4 GHz router because it delivered a higher Signal to Interference and Noise Ratio ($\mathtt{SINR}$), even when router load and resource differences between the bands often led to much higher performance in the 5 GHz band. Methods to increase capacity in heterogeneous networks, through biasing, blanking, and dynamic load balancing, are well studied \cite{andrews_overview_2014, jo_heterogeneous_2012, damnjanovic_survey_2011, ghosh_heterogeneous_2012}. 

Static biasing causes user equipment (UE) to connect to small cells if the received power from those cells is within the bias limit of the received power from the macro base station, reducing load on the macro tier. Recent work has sought to dynamically adjust these biases in a distributed and real-time manner, further increasing load balancing gains \cite{wang_load-aware_2014,ye_user_2013}. Furthermore, large potential gains in the uplink have been shown through decoupling uplink user association from the downlink through both analysis and detailed simulations\cite{smiljkovikj_analysis_2015,elshaer_downlink_2014, singh_joint_2015, boccardi_why_2015}. In \cite{smiljkovikj_analysis_2015}, stochastic geometry analysis is used to show that decoupling leads to significant uplink gain and an increase in fairness. Similar gains are shown in \cite{elshaer_downlink_2014} using a ray tracing prediction model. A minimum path loss association is shown to be optimal for the uplink in \cite{singh_joint_2015}. In both \cite{singh_joint_2015} and \cite{boccardi_why_2015}, uplink/downlink decoupling with biasing for the downlink is discussed. 

The existing analysis and simulation literature on biasing and decoupling uses single slope path loss models, though dual slope models have been shown to more closely match empirical results and to have significantly different characteristics in asymptotically dense networks \cite{erceg_urban/suburban_1992, hampton_urban_2006}. Multi-slope models more closely capture the relationship between the path loss and link distance in many cases. Some of these cases are summarized in \cite{zhang_downlink_2015}. Recent channel modeling activity in mmWave frequencies has also increased the interest in dual slope models because of significant blocking characteristics~\cite{rangan_millimeter-wave_2014, bai_coverage_2015, ghosh_millimeter-wave_2014}. Such models use different path loss exponents for line of sight (LOS) and non line of sight (NLOS) links. In \cite{zhang_downlink_2015}, coverage and capacity scaling is characterized with respect to the LOS path loss exponent. In~\cite{ding_performance_2015}, dual slope models are shown to affect coverage probability variation with density of the network. In~\cite{galiotto_effect_2014}, a similar effect is shown concerning area spectral efficiency and energy efficiency scaling with density. In this work, we extend the dual slope analysis to study user association techniques in heterogeneous networks.
  
\subsection{Contributions}
Static downlink biasing and pathloss based uplink association have been shown to boost throughput with single slope models~\cite{andrews_overview_2014, smiljkovikj_analysis_2015}. This paper develops insight on the performance of these techniques with dual slope models. We make the following observations, where we denote the path loss exponents below and beyond critical radius, and for single slope model  as $\alpha_0$, $\alpha_1$, and $\alpha$ respectively.
\begin{itemize}
\item Dual slope models lead to a much smaller gain from optimal downlink biasing compared to the single slope case, when $\alpha = \alpha_0$, though a higher overall rate. Dual slope models lead to steering of users to nearby small cells without biasing.
\item Biasing with both single and dual slope models increases the rate for the majority of users. However, with dual slope models the mismatch between bias for optimal median rate and that for optimal edge rate increases. This mismatch becomes more pronounced at moderate to high relative densities, where biasing for one leads to losses for the other.
\item Biasing rate gains for out-of-band small cells decrease as the relative density between the tiers increases. With single slope models, the gains are a much weaker negative function of the absolute densities of the tiers. With dual slope models, there is no such dependence on the absolute densities of the tiers.
\item The larger path loss exponent in a dual slope model determines biasing gain and the uplink/downlink decoupling gain. The larger path loss exponent is also much more significant than the smaller path loss exponent in determining the fraction of users associated to small cells.
\item Biasing optimally for the downlink captures most but not all of the gain from uplink/downlink decoupling. This trend is especially true with dual slope path loss models, as users are more likely to be associated with small cells with lesser (as compared to that for single slope) bias.\end{itemize}
Together, these observations underscore the importance of using dual or multi slope path loss models to analyze the performance of user association techniques.
\section{System Model}
A two-tier system is simulated, where the femto cells are out-of-band relative to the macro cells. UEs and BSs in each tier $i$ are placed according to a Poisson Point Processes (PPP) with density $\lambda_u$ and $\lambda_i$, respectively. A sample $p$ is drawn from a Poisson distribution with mean $\lambda$, and $p$ points from a 2-D uniform distribution are drawn on the grid $[-g, g]$. Users  associate  with BSs  according to a specified rule. Then, BS load, $\mathtt{SINR}$, and rate are calculated for the user at the origin. These values are measured over many experiments to generate distributions given each set of parameters. Without loss of generality, all measurements are made with respect to a fixed user at the origin. A full list of parameters is included in Table~\ref{tab:parameters}.

\index{commands!environments!table}%
\begin{table}
\caption [Simulation parameters and measurements.]{Simulation parameters and measurements.}
\label{tab:parameters}
\centering
\begin{tabular}{ | c | c |}
\hline
    \textbf{Parameter} & \textbf{Value}\\ \hline

    \hline
    Number of Tiers & 2 \\ \hline
    $\lambda_u$ & $200 \text{ UEs}/\text{km}^2$\\ \hline
    $\lambda_i$ & $10^{-1} \text{ to } 10^{2.5} \text{ BS}/\text{km}^2$\\ \hline
    $[-g, g]$   & $[-10, 10]$\\ \hline
    $h_{k,j}$   & iid exponential RV with $\mu = 1$\\ \hline
    $P_{t,i}$   & 46 dBm for Macro, 23 dBm for Femto\\ \hline
        $P_{t,u}$   & truncated channel inversion, maximum 20 dBm\\ \hline
    $\sigma^2$   & $-10$ dBm\\ \hline
    $\alpha$   & Single Slope Path Loss Exponent $2 \text{ and }3$\\ \hline

    $[\alpha_0$, $\alpha_1]$   & Path Loss Exponents $[2, 2], [2, 4], [3, 3],\text{ and }[3, 4]$\\ \hline
	$d_0$ & 100 m \\ \hline    
    $R_c$   & 30 m \\ \hline
    $B_i$   & $0 \text{ dB}\text{ to } 12 \text{ dB}$ \\ \hline
\end{tabular}
\end{table} 

\subsection{Propagation Model}
A Rayleigh Fading channel is used. A channel value $h_{k,j}$ between user $k$ and BS $j$ in tier $i$ is drawn from iid exponential distributions $ p_x (h_{k,j}, \lambda) = \lambda e^{-\lambda x} $, where $\lambda = 1$. Following~\cite{zhang_downlink_2015}, the received power with  single-slope model is 
\begin{equation}
\label{eqn:singleslope}
P_r = h_{k,j} P_t K \left(\frac{x_j}{d_0}\right)^{-\alpha}, 
\end{equation}
where $K$ encapsulates parameters such as antenna heights and gains, $x_j$ is the distance from  BS $j$, and that with  dual slope path loss model is
\begin{equation}
\label{eqn:dualslope}
P_r=\left\{
\begin{array}{c l}      
    h_{k,j} P_t K \left(\frac{x_j}{d_0}\right)^{-\alpha_0} & x \le R_c\\
    R_c^{\alpha_1 - \alpha_0} h_{k,j} P_t K \left(\frac{x_j}{d_0}\right)^{-\alpha_1} & x > R_c
\end{array}\right\},
\end{equation}
where $R_c$ is the critical radius, and  $R_c^{\alpha_1 - \alpha_0}$ factor is used for continuity purposes. 
\subsection{Downlink User Association}
For downlink association, two techniques are analyzed: highest received power and highest biased received power based association.
\label{sec:userassociation}
\subsubsection{Highest Received Power}
 A user $k$ associates with the BS $j$ that maximizes the $\mathtt{SNR}$. 
\begin{equation}
\label{eqn:highsinrdownlink}
\arg\max_{j} \{P_{r,{j}}  - \sigma^2 \},
\end{equation}
where $\sigma^2$ denotes the noise power. Note that such an association differs from association in real networks, where a UE can easily determine $\mathtt{SINR}$ from each BS using reference signals. $\mathtt{SNR}$ association is commonly used in simulation based studies due to intractability in computing $\mathtt{SINR}$ for all UEs \cite{smiljkovikj_analysis_2015, elshaer_downlink_2014}.
\subsubsection{Highest Biased Received Power}

A user $k$ is associated with the BS $j$ in tier $i$ that maximizes 
\begin{equation}
\max_{j} \{P_{r,{j}}  + B_i- \sigma^2\ \} ,
\end{equation}
where $B_i$ is the bias for the tier $i$.
This work assumes that all BSs within a tier share the same bias value, but  this value can be tuned for any given network configuration.
\subsection{Uplink User Association}
The strategies compared for uplink association are: coupled with the downlink, and decoupled with highest received power at the BS. 
\subsubsection{Coupled Association}
With this association, a user associates with the same BS on the uplink as in the downlink and selected based on one of the downlink association rules. 

\subsubsection{Highest Received Power at the Base Station (Uplink Pathloss Association)}
\label{sec:uplinkSNR}
 A user $k$ is associated with the BS $j$  that maximizes uplink received power, i.e. the BS index is
\begin{equation}
\arg\max_{j} \{ P_{r,j}(dB) - \sigma^2\},
\end{equation}
where
\begin{equation*}
P_{r,j} = \begin{cases}
 h_{k,j}KP_{t,u} \left(\frac{x_j}{d_0}\right)^{-\alpha_0} &  x \le R_c\\
 R_c^{\alpha_1 - \alpha_0} h_{k,j} P_{t,u} K \left(\frac{x_j}{d_0}\right)^{-\alpha_1}  & x > R_c.
\end{cases}
\end{equation*}
 Truncated channel inversion is used in the uplink. $P_{t,u}$ is the UE transmit power and includes truncated channel inversion, i.e., $P_{t,u}$ inverts the path loss up to the maximum transmitted power. Note that $P_{t,u}$ does not depend on tier to which the user is connected. Thus, this association rule is the same as transmitting to the BS that receives the highest instantaneous SNR from the user. 


\subsection{Simulation Setup}
\label{sec:performancemetrics}
For each parameter combination, measurements are collected and averaged over 2000 drops. For each drop, the BS locations, user locations, and the channel values are drawn from their respective distributions. The received power and $\mathtt{SINR}$ from BS $j$ in tier $i$ for each drop are computed as in (\ref{eqn:singleslope}) and (\ref{equation:finalSINRdownlink}). Note that in-cell interference is assumed to be perfectly nulled.
\begin{equation}
\label{equation:finalSINRdownlink}
\gamma_{downlink} = \frac{P_{r,{j}}}{\sum\limits_{l \neq j} P_{r,{l}} + \sigma^2}
\end{equation}
Full Channel State Information at the Transmitter (CSIT) is assumed with rate modulation. A user receives a downlink rate $r$ from BS $j$ in tier $i$, where 

\begin{equation}
\label{eqn:model_rate}
r=
    \frac{1}{N_{j}} \log_2(1 + \gamma_{downlink}) \text{ bps/Hz}
\end{equation}
and  $N_{j}$ is the number of associated users. This rate calculation translates to a system with fair, orthogonal resource partitioning with BS load $N_{j}$. Uplink rate is calculated similarly, with no intra-cell interference.
\section{Downlink Biasing Simulation Results}

\begin{figure}[t]
\centering
\includegraphics[width=3.4in]
{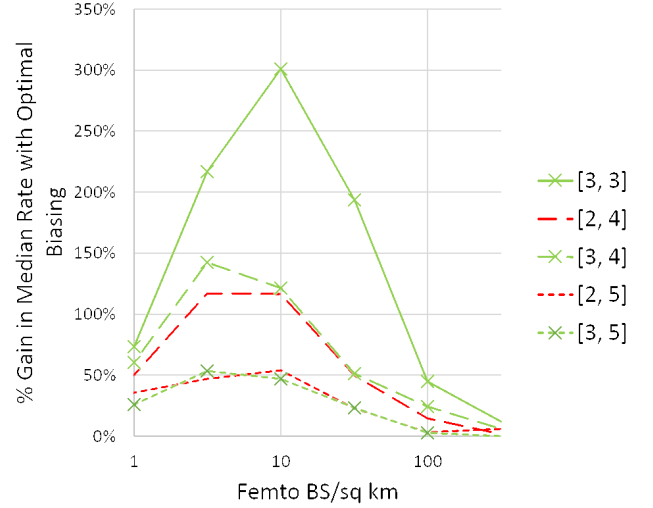}
\caption{Gain in downlink median rates with optimal biasing over no biasing for varying femto cell density and path loss exponents. Macro tier density is held constant at 1 BS/sq km.}
\label{fig:med50downlink}
\end{figure}
\begin{figure}[t]
\centering
\includegraphics[width=3.4in]
{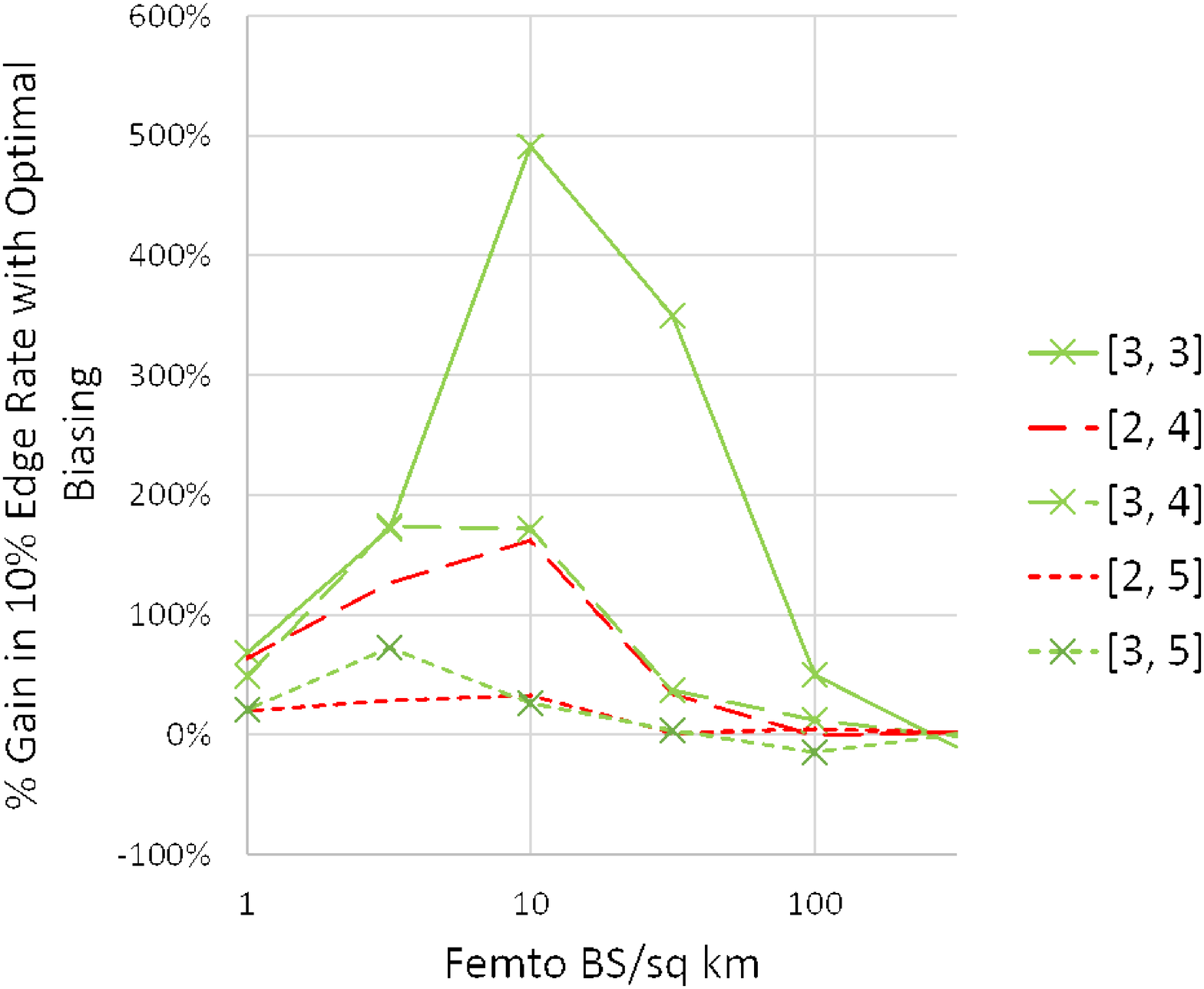}\caption{Gain in downlink edge rate with optimal biasing (for median rate) over no biasing for varying  femto cell density and path loss exponents. Macro tier density is held constant at 1 BS/sq km.}
\label{fig:med5downlink}
\end{figure}
\begin{figure}[t]
\centering
\includegraphics[width=3.4in]
{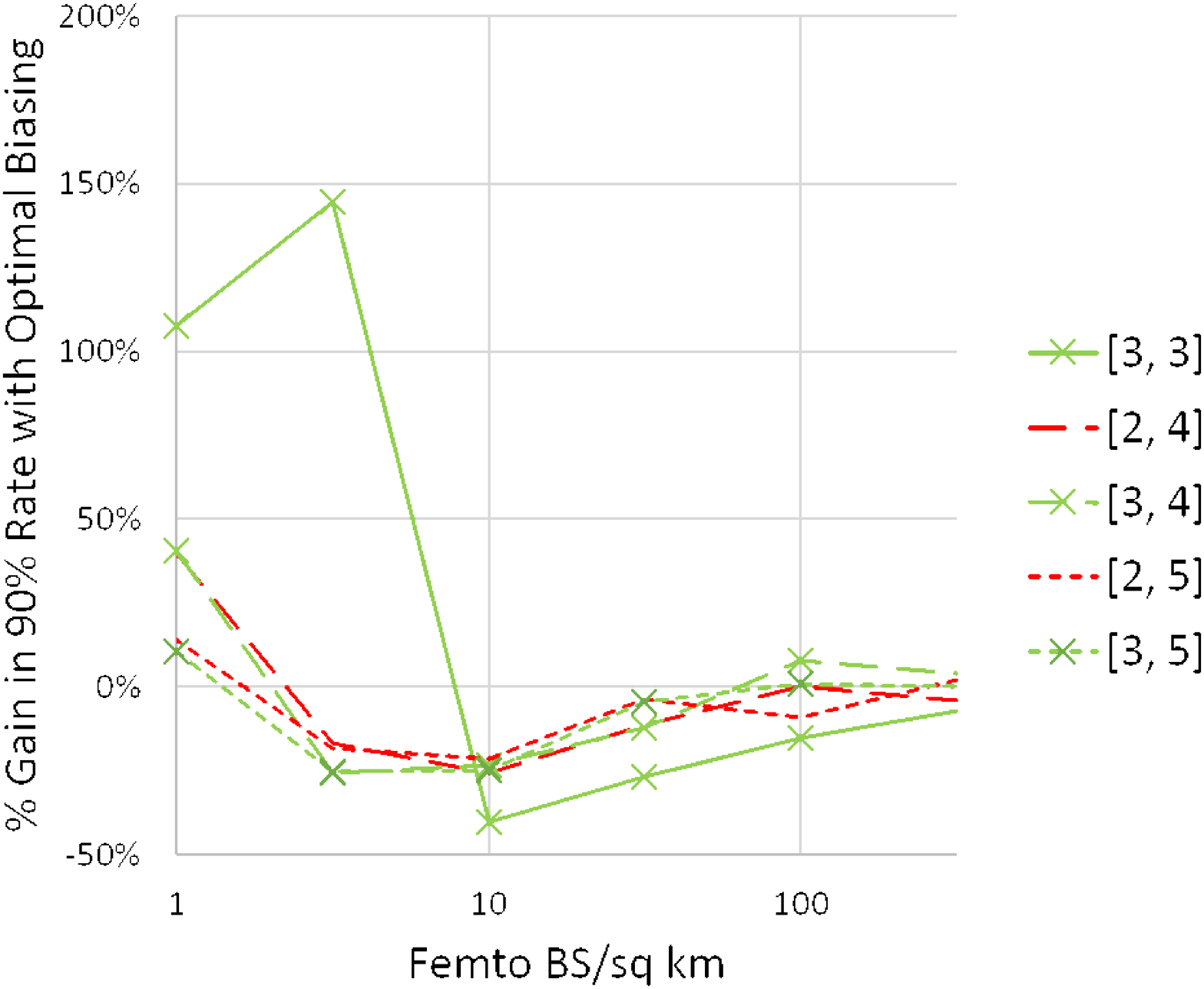}
\caption{Gain in downlink peak rate with optimal biasing (for median rate) over no biasing for varying  femto cell density and path loss exponents. Macro tier density is held constant at 1 BS/sq km.}
\label{fig:med90downlink}
\end{figure}
\label{sec:biasing}
\begin{figure}[t]
\centering
\includegraphics[width=3.4in]{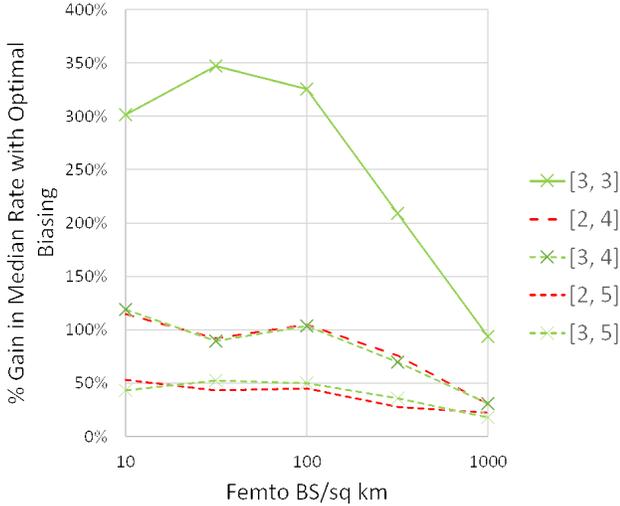}\caption{Optimal biasing gain for the median downlink user as the density of both tiers increases. Macro cell density is 10x less than the femto cell density. }
\label{fig:medianrawboth}
\end{figure}
In this section, the optimal biases are found as the bias values that maximize the median rate across drops for each parameter combination. The gain in median rate with optimal biasing over that with no biasing for various small cell densities and path loss exponents is shown in Fig.~\ref{fig:med50downlink}. Dual slope path loss model leads to a higher rate for the median user. However, as can be seen in the figure, there is a smaller additional gain from biasing when dual slope models are used. Using a dual slope path loss leads to a steering of users to nearby small cells without biasing; the larger transmit power of the macro station experiences a larger path loss if the BS is past the critical distance $R_c$. For large relative densities (when a small cell is likely to be within $R_c$), less biasing is required. 

Furthermore, with dual slope models the mismatch between the optimal bias for median rate and that for optimal edge ($10$th percentile) and peak ($90$th percentile) rate increases. 
Figs.~\ref{fig:med5downlink}~and~\ref{fig:med90downlink} show the gain in rate with optimal biasing for the  edge  and peak rate respectively. Under the stated assumptions, the biasing gain for the edge and median user persists even for large relative density with the single slope models for $\alpha = 3$. Single slope models lead to biasing that is beneficial to even the peak rate user at very low densities -- the load balancing gains from biasing are larger than the loss due to a reduced signal strength. For the dual slope cases, however, the edge and median users see a smaller gain. The edge user sees a much smaller gain with dual slope models than it would if the biasing were optimized for the edge user rather than the median user. Similarly, the peak rate user realizes biasing losses at any practical relative density with dual slope models. These high-rate users are already connected to small cells, and so they receive fewer resources as more users connect to small cells due to biasing. These losses are exacerbated if biasing values are chosen to maximize the edge rate instead of the median rate. Since biasing and techniques are often used to improve the rate for edge users, this tradeoff may be acceptable for the peak rate user but not the edge user. In settings with a higher difference between LOS and NLOS propagation, such as mmWave networks, this tradeoff may no longer be beneficial, and static biasing is less useful. The load balancing effects of the large NLOS path loss exponent may capture most of the gain, and biasing must be carefully tailored to increase the rates of target users. 

In~\cite{andrews_overview_2014}, bias factors are found to decrease with increasing small cell density for out-of-band small cells for single slope models. In this work, simulations show similar trends regarding optimal biases for dual slope models as well. Bias values are found to be negatively associated with relative density for out-of-band small cells, and biasing gains are also a function of the relative density of the tiers. Biasing gains do not strongly depend on absolute density of the small cells, just the relative density -- especially for the dual slope case. In Fig.~\ref{fig:medianrawboth}, the density of each tier increases while maintaining the relative density at a ratio of $10$ femto cells per macro cell, and the gains remain invariant for a large range of small cell density with dual slope models. There is a slight negative relationship, but it is much weaker than that in the relative density case. This result suggests that the number of small cells installed per macro BSs should be known for setting of the optimal biases. 

Finally, the figures demonstrate an non-intuitive result: NLOS propagation characteristics determine system performance. The gain in downlink rates for all users with biasing, is nearly identical for the $[2,4]\text{ and }[3,4]$ cases, and the $[2,5]\text{ and }[3,5]$ cases, respectively. The same trends can be found in the uplink/downlink decoupling results in the next section. The NLOS path loss exponent seems to determine both the interference from other cells and to which tier the UE will connect, and so it is important to use multi-slope models. Further analytical studies are required to validate these observations.

\section{Uplink/Downlink Decoupling Simulation Results}
\label{sec:dudesection}
\begin{figure}[t]
\centering
\includegraphics[width=3.4in]{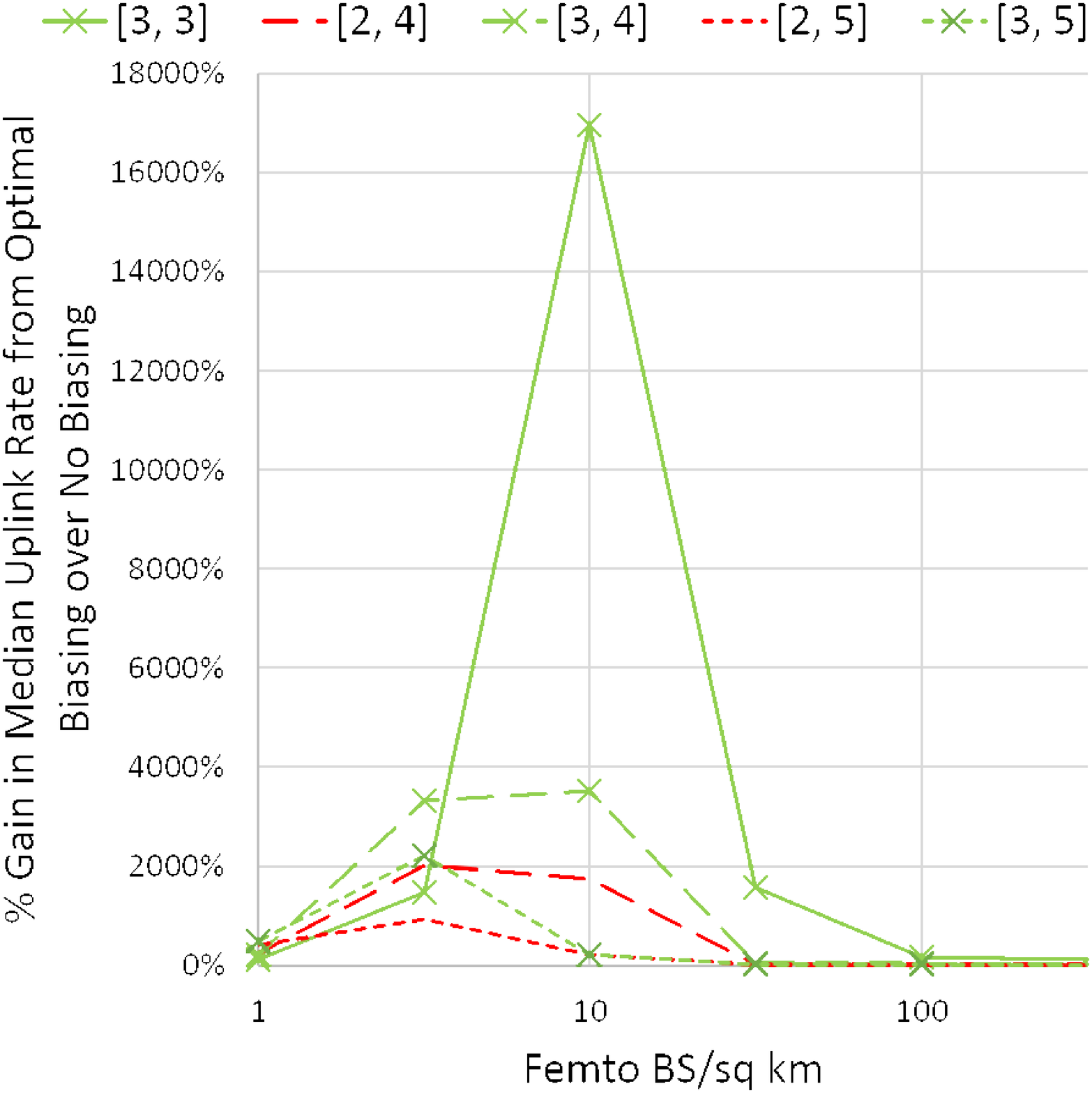}
\caption{Gain in median uplink rate for optimal downlink biasing when compared to downlink association without biasing. The macro tier density is held constant at 1 BS/sq km.}
\label{fig:withoutbiasing}
\end{figure}
\begin{figure}[t]
\centering
\includegraphics[width=3.4in]
{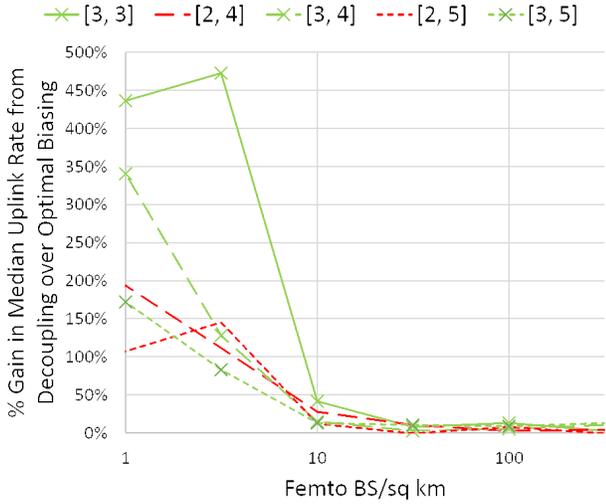}\caption{Gain in median uplink rate due to decoupling when compared to downlink association with optimal downlink biasing. The macro tier density is held constant at 1 BS/sq km.}
\label{fig:withbiasing}
\end{figure}

In~\cite{singh_joint_2015} it is shown that with coupled association, biasing for load balancing in the downlink would also benefit the uplink, as  users are more likely to connect to closer, low-power BSs that also exhibit a smaller uplink path loss. However, it is shown that in \cite{singh_joint_2015,boccardi_why_2015}  that the decoupled uplink would still outperform optimal coupled association. In this work, these claims are further investigated in the context of dual slope model.  
Fig.~\ref{fig:withoutbiasing} shows the gain in median uplink rate due to optimal downlink biasing over the no biasing case. Note that, as defined, optimal biasing only maximizes downlink median rate, rather than the joint downlink-uplink rate. Thus, optimal biasing that optimizes joint uplink and downlink capacity could yield further gains as shown in \cite{singh_joint_2015}. Fig.~\ref{fig:withbiasing} shows the gain in median user uplink rate from decoupling when compared to coupled association with optimal downlink biasing. The majority of the gain in uplink rate is captured by optimal biasing for the downlink, for both single slope and dual slope models. However, under our assumptions, decoupling still yields gains over optimal downlink biasing up to a relative density of 10 small cells per macro cell. This gain is seen to be lower for dual slope models. 

\begin{figure}[t]
\centering
\includegraphics[width=3.4in]{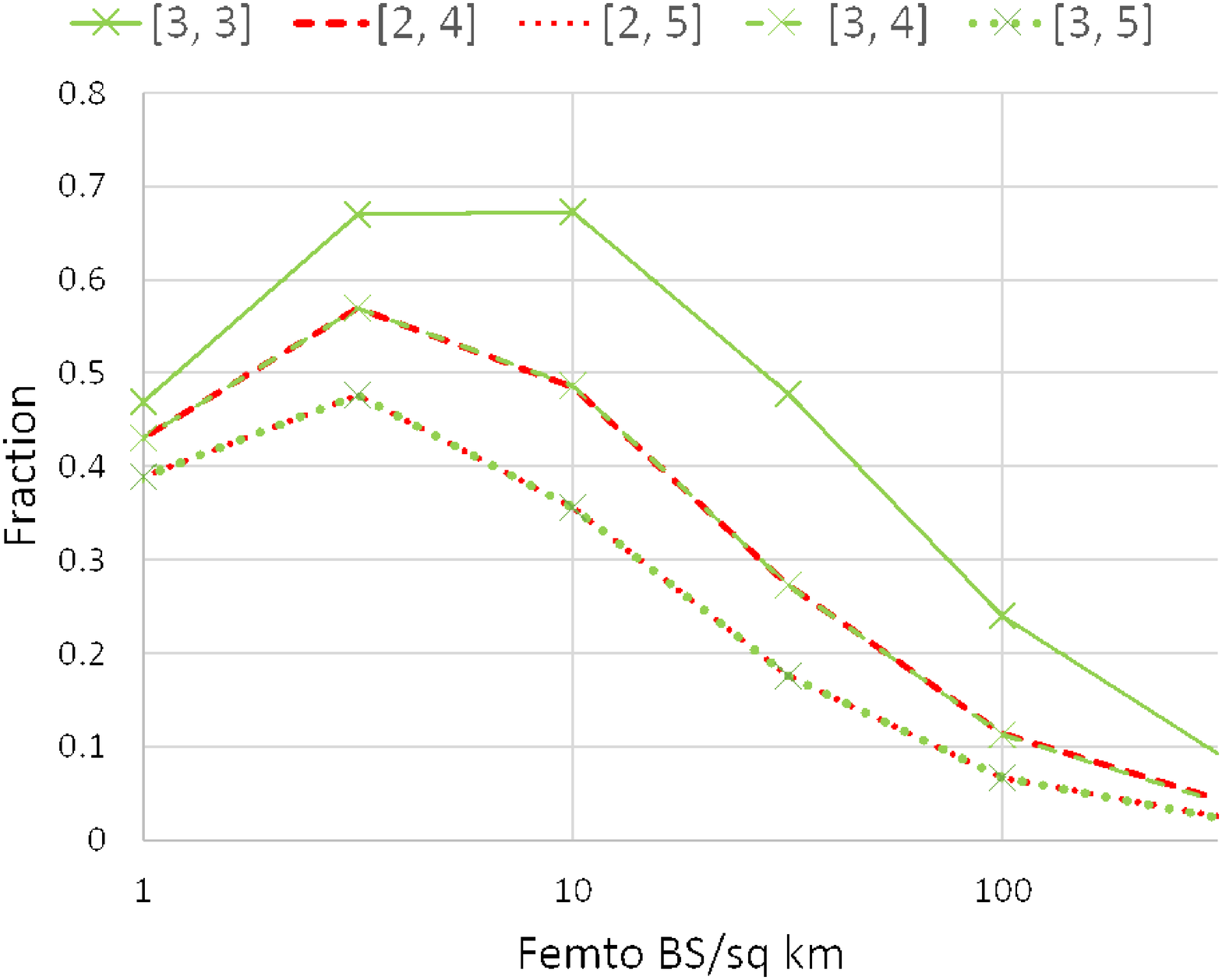}
\caption{Fraction of users associated differently for uplink and downlink when path loss association is used for the uplink and downlink $\mathtt{SNR}$ is used for the downlink. The macro tier density is held constant at 1 BS/sq km.}
\label{fig:fractionassociatediff}
\end{figure}

\begin{figure}[t]
\centering
\includegraphics[width=3.4in]{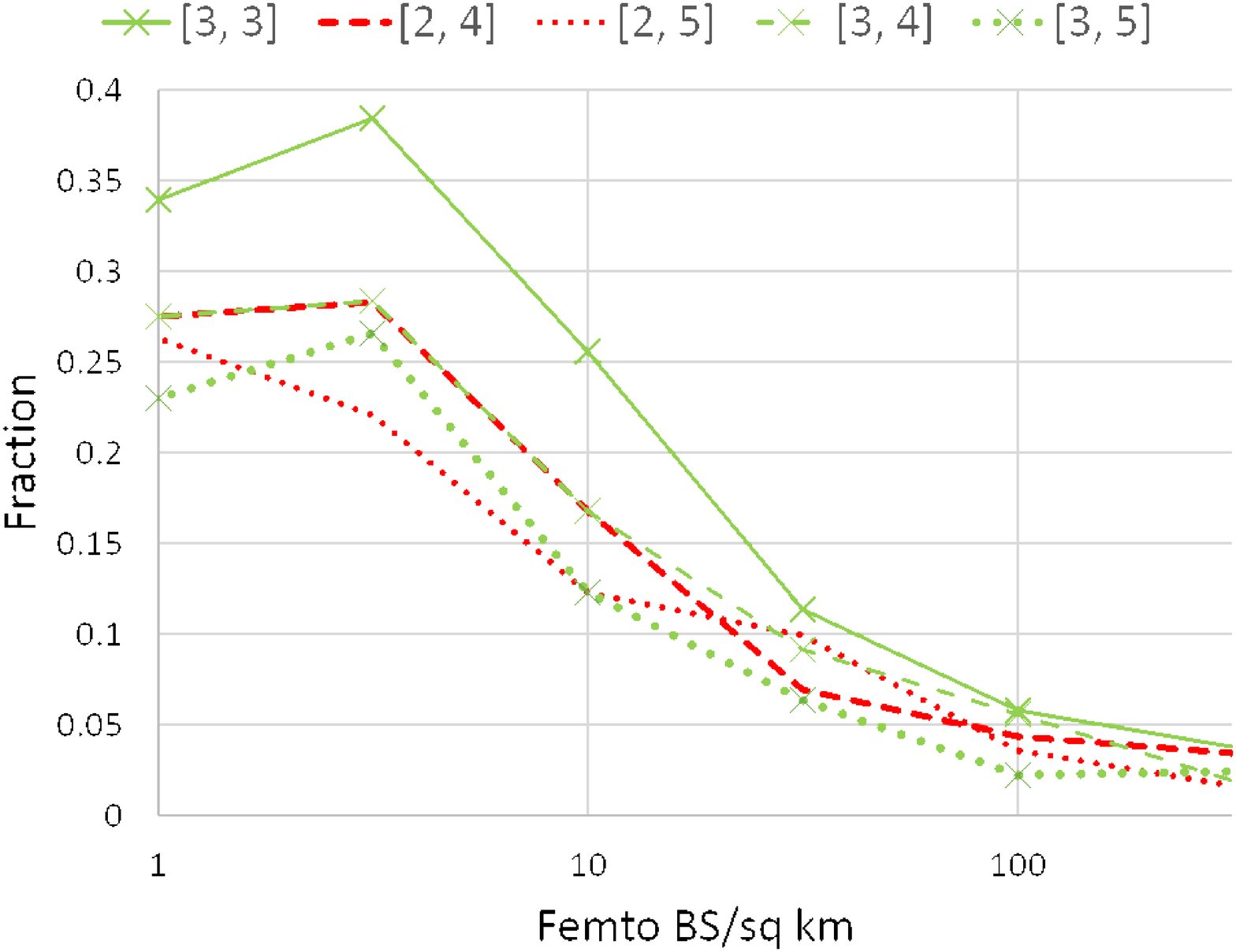}
\caption{Fraction of users associated differently for uplink and downlink, when path loss association is used for the uplink and optimal downlink biasing is used for the downlink. The macro tier density is held constant at 1 BS/sq km.}
\label{fig:fractionassociatediff_optimalbiasing}
\end{figure}

These results can largely be explained by the fraction of users associated to the small cells in each of the cases. Figs.~\ref{fig:fractionassociatediff}~and~\ref{fig:fractionassociatediff_optimalbiasing} show the fraction of users that are connected to different BSs for the uplink than they are for the downlink, when path loss association is used for the uplink. Fig.~\ref{fig:fractionassociatediff} shows the case without downlink biasing, and Fig.~\ref{fig:fractionassociatediff_optimalbiasing} shows the case with downlink biasing. Downlink biasing reduces the number of users associated sub-optimally for the uplink by a factor of 2. The use of dual slope path loss models further reduces this number. These figures also show the effect of the larger, NLOS exponent, as discussed in the previous section. The larger path loss exponent largely determines the fraction of users associated to each tier. 


\section{Conclusion}
This work demonstrates the importance of using dual slope path loss models to analyze user association schemes in heterogeneous networks. Dual slope models lead to lower biasing downlink rate gains and uplink/downlink decoupling uplink rate gains. For many practical relative densities between the macro and the small cell tier, it may not be beneficial to decouple uplink/downlink if biasing is already used for downlink load balancing. We also observe the large relative significance of the larger path loss exponent in biasing gain, uplink/downlink decoupling gain, and in the fraction of users associated to small cells. These results add to an increasing call to use dual or multi slope models instead of the idealized, single slope path loss model. Dual slope models better match real deployments and, as demonstrated in this paper, often lead to different results when analyzing techniques. Future work should analyze cases in which different tiers have different propagation characteristics, and when the small cells are in-band to the macro cells.





\bibliographystyle{IEEEtran}
\bibliography{IEEEabrv,bibliography2}
%

\end{document}